\documentclass[a4]{article}

\usepackage{stfloats}
\usepackage{mathrsfs}
\usepackage{anysize}
\marginsize{3cm}{2cm}{1cm}{1cm}

\date{}

\makeatletter

\newcommand{\JLA}[1]{\begingroup\color{black}#1\endgroup}
\newcommand{\DM}[1]{\begingroup\color{black}#1\endgroup}

\usepackage{multicol}
\usepackage[usenames,dvipsnames]{xcolor}
\usepackage{subfig}
\usepackage{caption}
\newcommand{\incFig}[1]{%
\includegraphics{fig_#1}%
}

\usepackage{mathtools, amsmath, amssymb}
\newcommand\CPU{Intel\textregistered\ Core\texttrademark\ i5-2500K CPU (3.30GHz)}

\DeclareMathOperator{\img}{img}

\newcommand{\stratStyle}[1]{\texttt{\textbf{#1}}}
\DeclareMathOperator{\stratAltRuleApp}{\stratStyle{altRuleApp}}
\DeclareMathOperator{\stratParallel}{\stratStyle{parallel}}
\DeclareMathOperator{\stratRepeat}{\stratStyle{repeat}}
\DeclareMathOperator{\stratRevive}{\stratStyle{revive}}
\DeclareMathOperator{\stratLeftPredicate}{\stratStyle{leftPredicate}}
\DeclareMathOperator{\stratRightPredicate}{\stratStyle{rightPredicate}}
\DeclareMathOperator{\stratFilterSubset}{\stratStyle{filterSubset}}
\DeclareMathOperator{\stratFilterUniverse}{\stratStyle{filterUniverse}}
\DeclareMathOperator{\stratSortSubset}{\stratStyle{sortSubset}}
\DeclareMathOperator{\stratSortUniverse}{\stratStyle{sortUniverse}}
\DeclareMathOperator{\stableSort}{\stratStyle{stableSort}}
\DeclareMathOperator{\stratTakeSubset}{\stratStyle{takeSubset}}
\DeclareMathOperator{\stratTakeUniverse}{\stratStyle{takeUniverse}}
\DeclareMathOperator{\stratAddSubset}{\stratStyle{addSubset}}
\DeclareMathOperator{\stratAddUniverse}{\stratStyle{addUniverse}}

\makeatother

\begin{document}%
\title{Generic Strategies for\\ Chemical Space Exploration}
\maketitle
\setcounter{page}{1}

\author{
\begin{center}{\large Jakob~L.~Andersen$^{1,2}$,~Christoph~Flamm$^2$,~Daniel~Merkle$^{1}$,~and~Peter~F.~Stadler$^{2-7}$}\\[1.5cm]
$^1$ Department of Mathematics and Computer Science\\ University of Southern Denmark, Denmark\\[0.2cm]
$^2$ Institute for Theoretical Chemistry, University of Vienna, Austria.\\[0.2cm]
$^3$ Bioinformatics Group, Department of Computer Science, and\\
Interdisciplinary Center for Bioinformatics, University of Leipzig, Germany.\\[0.2cm]
$^4$ Max Planck Institute for Mathematics in the Sciences, Leipzig, Germany.\\[0.2cm]
$^5$ Fraunhofer Institute for Cell Therapy and Immunology, Leipzig, Germany.\\[0.2cm]
$^6$ Center for non-coding RNA in Technology and Health\\University of Copenhagen, Denmark.\\[0.2cm]
$^7$ Santa Fe Institute, 1399 Hyde Park Rd, Santa Fe, NM 87501, USA\end{center}}

\begin{abstract}
  Computational approaches to exploring ``chemical universes'', i.e., very
  large sets, potentially infinite sets of compounds that can be
  constructed by a prescribed collection of reaction mechanisms, in
  practice suffer from a combinatorial explosion. It quickly becomes
  impossible to test, for all pairs of compounds in a rapidly growing
  network, whether they can react with each other. More sophisticated and
  efficient strategies are therefore required to construct very large
  chemical reaction networks.

  Undirected labeled graphs and graph rewriting are natural models of
  chemical compounds and chemical reactions. Borrowing the idea of partial
  evaluation from functional programming, we introduce partial applications
  of rewrite rules. Binding substrate to rules increases the number of
  rules but drastically prunes the substrate sets to which it might match,
  resulting in dramatically reduced resource requirements.  At the same
  time, exploration strategies can be guided, e.g.\ based on restrictions
  on the product molecules to avoid the explicit enumeration of very
  unlikely compounds. To this end we introduce here a generic framework for
  the specification of exploration strategies in graph-rewriting
  systems. Using key examples of complex chemical networks from sugar
  chemistry and the realm of metabolic networks we demonstrate the
  feasibility of a high-level strategy framework.

  Graph grammars in conjunction with efficient, versatile exploration
  strategies are a powerful framework for combinatorial chemistry
  applications, allowing detailed investigations in very large
  chemical spaces, \DM{which is the foundation for understanding
    function of biological systems}. The ideas presented here can not only be used 
for a strategy-based chemical space exploration that has close correspondence of experimental results, but
  are much more general. In particular, the framework can be used
  to emulate higher-level transformation models such as illustrated in
  a small puzzle game.
\end{abstract}

\section{Introduction}

The systematic computational exploration of chemical spaces have become
topic of high practical relevance e.g.\ in drug design
\cite{Eberhardt:11,Dow12,Reymond:12,Wong:12}. Recent efforts to gain
insights into the distribution of properties in chemical spaces include the
construction of large databases of hypothetical compounds. The ``chemical
universe database GDB-17'' \cite{Ruddigkeit:17}, for instance comprises
166.4 billion molecules of up to 17 atoms of C, N, O, S, and halogens
covering the size and composition range of typical lead compounds. Beyond
the diversity of molecules and their properties, however, potential
synthesis pathways leading to them are a crucially important consideration
in practice. This calls for methods to systematically explore chemical
spaces in terms of restricted types of chemical reactions. Here we demonstrate
how this can be achieved in a natural way by means of graph grammars in
conjunction with efficient exploration strategies.

The \emph{structural formula} of a chemical compound is a graph that
represents the connectivity and mutual arrangements of its atoms.  Atom
types are given as vertex labels, while edges represent bond types.  At
this level of modeling, chemical reactions are naturally represented as
graph transformations.  Chemical reactions are explained and categorized in
terms of \emph{reaction mechanisms} that encapsulate the local changes of
chemical bonds.  In the formal framework of graph grammars, reaction
mechanisms correspond to the productions (rules).  Because of this
conceptual alignment between chemistry and graph grammars, a variety of
artificial chemistry models of different degree of chemical realism have
been devised on this basis \cite{Dittrich:01}.  Of course, these purely
combinatorial models of chemistry have their limitations.  Deliberately
disregarding the spatial embedding of molecules they cannot capture many
aspects of stereochemistry and they are restricted to (over)simplified
models of reactions energies and reaction kinetics.  Graph grammar models
are nevertheless of practical interest when the task is to explore large
areas of chemical spaces and they provide a means of analyzing regularities
in very large reaction networks.

Several graph rewriting tools have become available in the recent past, see
\cite{Fern10} for an overview. Application areas beyond chemistry include
model checking and verification, proof representation, and modeling control
flow of programs among many others. A strategy language to control the
application of graph rewriting rules has been presented in
\cite{Kirchner11} for PORGY \cite{Andrei11,Pinaud12}. A strategy framework
for exploring chemical spaces has very different design goals from the
properties desirable for applications within different areas of computer
science. For instance, chemical graph grammar rule frequently merge or
split graphs, since connected components correspond to individual
molecules. Hence a chemically motivated component handling is required. The
theoretically reachable chemical spaces can be infinite, e.g., when the
rules and the starting material allow polymers to form. Exploration thus
may not halt except due an enforced resource (size) limitation.  Decisions
on how to expand the space are usually heavily influenced by chemical
properties or additional data sources. Furthermore, the goal of an analysis
might also be motivated by a chemical question such as the detection of
chemical subspaces or the need to find specific chemical transformation
patterns.

In this paper different types of ``chemistries'' will be used to demonstrate the
different aspects of the strategic construction framework for fractions of
the chemical space. The Diels-Alder reaction \cite{Diels:1928}, a   sigmatrope cycloaddition
reaction between a conjugated diene and an   alkene will serve as an example
for a combinatorial complex chemical   space emerging from a single reaction
rule. The formose reaction   \cite{Butlerov:1861}, which subsumes the
formation of sugars from   formaldehyde is used to explore the impact of
changes in ``chemistry''   (i.e.\ the set of reaction rules) on the structure
and complexity of the   chemical space. HCN chemistry is used to
illustrate how construction of chemical space can be biased with
experimental data. 

The outline of the paper is as follows. In the section ``Formal Framework''
we will present the underlying framework of graph transformation including the Double Pushout
Approach. Secondly, in ``Transformation by Partial Rule Application'', we describe a method for efficient
calculation of rule application.
Thirdly, general strategies to explore chemical spaces will be introduced in the ``Strategies'' section.
In the ``Results'' section we will apply our framework to several before-mentioned complex chemical
settings. Finally a puzzle game will be used to illustrate the generality of our approaches.

\section{Formal Framework}
\subsection{Chemical Graph Rewriting with the Double Pushout Approach}
Molecules are always represented by connected graphs.  Chemical reactions,
however, more often than not, involve two or more interacting molecules as
their ``input'' (educts) and there is no guarantee that the ``output''
(products) is connected.  Thus we have to consider graph transformations
that operate on not necessarily connected graphs.  More precisely, we
regard a graph $G$ here as a multiset $\{g_1, g_2,\dots, g_{\#G}\}$ of its
$\#G$ connected components.  All graphs are simple, i.e., without loops and
parallel edges.  Double and triple bonds are viewed as edge labels rather
than multiple edges.

Several abstract formalisms for graph transformation have been explored in
the literature, see e.g.,~\cite{handbook1} for a detailed introduction.  We
found that the so-called Double Pushout (DPO) approach provides the most
intuitive direct encoding of chemical reactions and the closest connection
to the language of chemistry.  A DPO transformation rule $p
=(L\xleftarrow{l} K \xrightarrow{r} R)$ consists of three graphs $L$, $R$
and $K$ known as the left, right and context graph, respectively, and two
graph morphisms $l$ and $r$ that determine how the context is embedded in
the left and the right graph.  The rule $p$ can be applied to a graph $G$
if the left graph $L$ can be found in $G$ and some additional consistency
conditions are satisfied.  This is modeled by the requirement that there is
a \emph{matching morphism} $m : L\to G$ that describe how $L$ is contained
in $G$.  Intuitively, the copy of $L$ is replaced within $G$ by $R$ in such
a way that the context $K$ is left intact, resulting in the transformed
graph $H$.  This operation, the derivation $G\xRightarrow{p, m} H$, is
described in the framework of category theory by the requirement that the
following commutative diagram exists:
\begin{align}
\begin{minipage}{0.9\textwidth}
\centering
\incFig{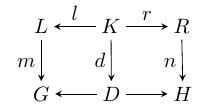}
\end{minipage}\label{eq:deriv}
\end{align}
The derivation $G\xRightarrow{p, m} H$ implicitly define the intermediary
graph $D$ and the result graph $H$ as well as morphisms $d : K\to D$ and $n
: R\to H$ that fix how the context and the right graph of the rule are
embedded in the intermediary and the result graph, respectively.  In terms
of molecules (connected components) we can write $\{g_1, g_2, \dots,
g_{\#G}\}\Rightarrow \{h_1, h_2,\dots,h_{\#H}\}$.

\begin{figure}
\centering
\incFig{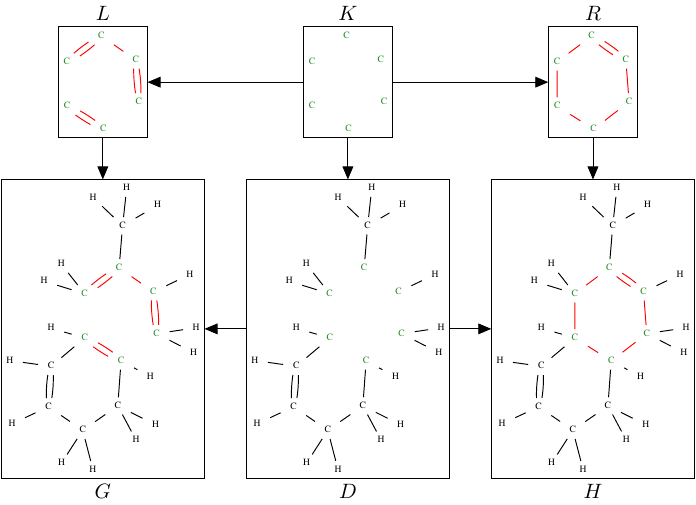}
    \caption[Example of Chemical Derivation]
	{Example of a chemical derivation from cyclohexadiene and
      isoprene using a Diels-Alder transformation.  The edges changed by
      the transformation is shown in red and the vertices from $K$ are
      shown in green.  Note that edges shown in parallel are in the
      underlying graphs a single edge with a special label to encode a
      specific chemical bond.  }
\label{fig:dpoDielsAlder}
\end{figure}

In applications to modeling chemistry, several additional requirements must
be satisfied.  Conservation of mass and atom types dictates that the
restrictions of $r$ and $l$ to the vertex sets (atoms) are bijective.
Furthermore, $m$ (and by extension $d$ and $n$) are subgraph isomorphisms
and hence injective.  We note in passing that this guarantees the existence
of a bijection $a : V(G)\rightarrow V(H)$ known as the \emph{atom mapping}.
In the DPO formalism, furthermore, the existence of an inverse production
$p^{-1} = (L\xleftarrow{l} K \xrightarrow{r} R)$, corresponding to the
reverse chemical reaction, is guaranteed. Some more basic properties of
chemical graph grammars can be found in \cite{Andersen:13r}.
Fig.~\ref{fig:dpoDielsAlder} shows an example of a chemical derivation.

\subsection{Proper Derivations}
Consider a valid derivation $\{g_1, g_2\}\xRightarrow{p, m}\{h_1, h_2\}$
and an arbitrary graph $g'$.  Clearly, the derivation $\{g_1, g_2, g'\}\xRightarrow{p,
  m}\{h_1, h_2, g'\}$ is also valid because the images of $m$ and $n$ are
contained in $\{g_1, g_2\}$ and $\{h_1, h_2\}$, respectively.  The graph
$g'$ is irrelevant for the transformation.  We call a derivation
$G\xRightarrow{p, m}H$ \emph{proper} if $\img{m}\cap g_i\ne\emptyset$ for
all $g_i\in G$. It is not hard to see that the inverse of a proper
derivation is again proper. 
Throughout the following sections we will assume every derivation to be
proper, unless otherwise stated.

\subsection{Derivation Graphs}
Chemical reaction networks can be represented as directed
(multi)hypergraphs whose vertices are the molecules of the ``chemical
universe'' under consideration and whose hyperedges represent chemical
reactions \cite{Zeigarnik:00a}.  Here, it is important to consider
hyperedges as multisets to accommodate the stoichiometric coefficients,
i.e., the multiplicities in which molecules enter a chemical reaction such
as $2H_2 + O_2 \to 2H_2O$.  Such networks can be constructed from
experimentally observed data.  An example is the Network of Organic
Chemistry (NOC) \cite{Bishop:2006,Fialkowski:2005,Grzybowski:2009}, which
shows a non-trivial organization concentrated around a core region of about
300 synthetically important building blocks and industrial compounds.
Metabolic networks consist of the enzymatically catalyzed reactions
constituting the chemical basis of modern life forms.  They are available
from dedicated databases, see e.g.,~\cite{Karp:11}.

\begin{figure}
\centering
\subfloat[]{
\incFig{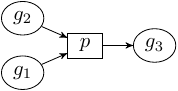}
\label{fig:dg:general}
}\qquad
\subfloat[]{
\incFig{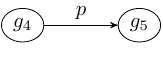}
\label{fig:dg:iso}
}\qquad
\subfloat[]{
\incFig{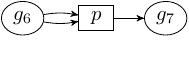}
\label{fig:dg:multi}
}
\caption[Visualization of Derivation Graphs]
	{A bipartite graph notation for directed (multi-)hypergraphs, in which the production itself is
  drawn as a special type of intermediate vertex, is used in most cases;
  \subref{fig:dg:general}, $\{g_1, g_2\}\xRightarrow{p} g_3$.  We only make
  an exception for 1-to-1 transformations (isomerization reactions);
  \subref{fig:dg:iso}, $g_4\xRightarrow{p} g_5$.  Multiplicities are
  indicated by multiple arcs; \subref{fig:dg:multi}, $\{g_6,
  g_6\}\xRightarrow{p} g_7$.  }
\label{fig:dg}
\end{figure}

In the framework of graph grammar models, an analogous \emph{derivation
  graph} can be defined.  Its vertex set consists of the connected labeled
graphs $\mathfrak{G}$ that represent the molecules.  Directed hyperedges
connect the multisets $G\subseteq \mathfrak{G}$ and $H\subseteq
\mathfrak{G}$ only if there is a proper derivation $G\xRightarrow{p,m}H$.
The conventions for visualizing hyperedges adhere to the three examples in
Fig.~\ref{fig:dg}.

\section{Transformation by Partial Rule Application}
The core strategy to expand the underlying derivation graph is the
discovery of new graphs by means of proper derivations implied by the direct
application of rules. Given a rule $p = (L\leftarrow K\rightarrow R)$ and a
set of graphs $\mathfrak{U}$, the task is to find all proper derivations
$G\xRightarrow{p} H, G\subseteq \mathfrak{U}$ where $G$ and $H$ are
multisets of graphs.  This can be done by a testing of all $k$-multisubsets
of $\mathfrak{U}$ for all $1\leq k\leq \#L$.  Since nearly all chemical
reactions are mono-molecular or bi-molecular, we can restrict ourselves to
$\#L \leq 2$, at least when elementary reactions are of primary interest.
Still, the number of multisets is $O(|\mathfrak{U}|^2)$.  In the worst
case, all unique multisets may give successful transformations, often
leading to a combinatorial explosion that quickly becomes unmanageable.  In
the following section we show that a more detailed control of the multisets
that are considered for transformation is desirable.

The key concept is partial rule composition~\cite{Andersen:13r}, i.e., the
binding of graphs to rules, resulting in partial rules that can be applied
more efficiently in an exploration strategy. The idea is analogous to
partial evaluation of functions by binding some of the variables. Full
graph transformations are computed as repeated partial rule application in
this framework. For the sake of brevity, we only sketch the idea here and 
omit a complete formal definition of partial rules. 

\begin{figure}
\centering
\subfloat[]{
\incFig{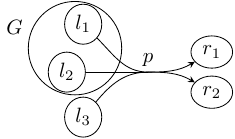}
%
\label{fig:partialRuleAppPre}
}\quad
\subfloat[]{
\incFig{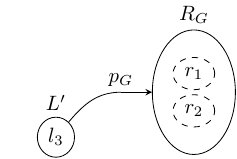}
%
%
\label{fig:partialRuleAppPost}
}\quad
\subfloat[]{
\incFig{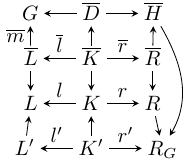}
%
\label{fig:partialRuleAppDiagram}
}
\caption[Partial Rule Application]
	{Partial application of some rule $p = (L\xleftarrow{l}
  K\xrightarrow{r} R)$ to a graph $G$, with $L = \{l_1, l_2, l_3\}$ and $R
  = \{r_1, r_2\}$.  The partial application is done through a partial
  matching morphism $\overline{m} : \overline{L}\rightarrow G$ with
  $\overline{L} = \{l_1, l_2\}$.  The application results in a new rule,
  $p_G = (L'\xleftarrow{l'} K'\xrightarrow{r'} R_G)$ with $L' = \{l_3\}$,
  for which $R$ is a subgraph of $R_G$.  The transformed graph of $G$,
  called $\overline{H}$, is also a subgraph of $R_G$.
  Fig.~\subref{fig:partialRuleAppDiagram} is the diagram of subgraph
  relations for a general partial rule application.}
\label{fig:partialRuleAppEx}
\end{figure}

A partial rule application of a rule $p=(L\xleftarrow{l}K\xrightarrow{r}R)$
with $L = \{l_1, l_2, \dots, l_{\#L}\}$ to a graph $G$, is a generalization
of a full transformation of $G$ in which only some but not all components
of $L$ do not match $G$. Thus $L$ is partitioned into the matching part
$\overline{L}\ne\emptyset$ and the non-matching remainder $L'$. The
restriction $\overline{l}$ of $l:K\to L$ to the pre-image $\overline{K}$ of
$\overline{L}$ defines the partial transformation rule $\overline{p} =
(\overline{L}\xleftarrow{\overline{l}}\overline{K}\xrightarrow{\overline{r}}\overline{R})$.
Using the restricted matching morphism $\overline{m} : \overline{L}
\rightarrow G$ it can be applied to $G$ resulting in graph $\overline{H}$.
The remainder $L'$ of $L$ gives rise to a new rule $p_G =
(L'\xleftarrow{l'} K'\xrightarrow{r'} R_G)$ whose right graph consists of
the transformed version of $G$ as well the original right graph $R$, i.e.,
it contains both $\overline{H}$ and $R$ as subgraphs. A formal,
diagrammatic representation is given in
Fig.~\ref{fig:partialRuleAppDiagram}.  An abstract partial application is
shown in Fig.~\ref{fig:partialRuleAppPre} and \ref{fig:partialRuleAppPost}.

Given a not necessarily connected graph $G$ and DPO transformation rule $p
= (L\xleftarrow{l} K\xrightarrow{r} R)$, our task is to construct all 
partial rules obtainable by binding $G$ to $p$. These partial rules can
then be applied to further graphs, allowing for more efficient exploration
strategies. The following algorithm enumerates these partial rules:

\begin{enumerate}
\item For all $l_i\in L$ find the set of all subgraph isomorphisms of $l_i$
  to $G$.  That is, find $M_i = \{m \mid m : l_i\rightarrow G\text{ is a
    subgraph isomorphism}\}$ for $1\leq i \leq \#L$.
\item For all nonempty subsets $\overline{L}$ of $L$, construct all partial
  matching morphisms, $\overline{m}$, by merging morphisms from each $M_j,
  l_j\in \overline{L}$.  Note, that each $\overline{m}$ must be injective.
\item For each partial matching morphism, $\overline{m}$, apply $p$ to $G$
  with $\overline{m}$ to obtain a new rule $p_G = (L'\xleftarrow{l'}
  K'\xrightarrow{r'} R_G)$.
\end{enumerate}
The partial matching morphisms constructed from considering $\overline{L} =
L$ are actually full matching morphisms, and so the resulting rule has $L'
= K' = \emptyset$.  In this case $p_G$ represents the creation of $R_G$
from an empty graph, and $G\xRightarrow{p, \overline{m}} R_G$ is a valid
derivation.  If $G$ is connected, the derivation will additionally be
proper.

In the following section we will regard a rule $p$ as a function on sets
of graphs, defined provisionally as:
\begin{align*}
  p(\mathfrak{U}) &= \mathfrak{U}\cup \bigcup_{\substack{G\xRightarrow{p}
      H\\ G\subseteq \mathfrak{U}}} H
\end{align*}
That is, the result of applying $p$ to a set of graphs, $\mathfrak{U}$, is
$\mathfrak{U}$ itself along with all graphs derivable from $\mathfrak{U}$ 
using $p$.

\subsection{Complex Graph States}
Consider the problem of applying a rule $p$ twice to a set of graphs $\mathfrak{U}$.
That is, finding $\mathfrak{U}_2 = p(\mathfrak{U}_1)$ for $\mathfrak{U}_1 = p(\mathfrak{U})$.
By our definition of rule application we have $\mathfrak{U}\subseteq \mathfrak{U}_1$, so when the algorithm described above is used for evaluating $p(\mathfrak{U}_1)$ it will find not only new derivations but also all derivations found when evaluating $p(\mathfrak{U})$.
We therefore use a more complex state than simply sets of graphs.
A \emph{graph state} $F$ is defined as a pair of ordered sets of graphs $(\mathcal{U}, \mathcal{S})$ with $\mathcal{S}\subseteq \mathcal{U}$.
The elements, $\mathcal{U}$ and $\mathcal{S}$, will be referred to also as  $U(F)$ and $S(F)$ respectively, where $U$ and $S$ are functions on the graph state.
In the following we will denote $U(F)$ as the \emph{universe} of the graph state $F$ and $S(F)$ as the \emph{subset} of the state.
The order of graphs in the subset and in the universe is independent and is arbitrary unless otherwise stated.

We define the  application of a rule $p$ to a graph state $F$ in the following manner.
Let $H'$ be all connected graphs derivable from $U(F)$  with $p$ such that at least one graph from $S(U)$ is being transformed in each derivation:
\begin{align}
	H' &= \{h\in H\mid G\xRightarrow{p} H : G\subseteq U(F) \wedge G\cap S(F) \neq \emptyset\}	\label{eq:ruleDef}
\end{align}
The result $F' = p(F)$ is such that
\begin{align}
	U(F') &= U(F)\cup 	H'			\label{eq:ruleUniverseDef}					\\
	S(F') &= H'\backslash U(F)										\label{eq:ruleSubsetDef}
\end{align}
That is, the resulting universe contains the input universe and all derived graphs, and the resulting subset contains all new graphs which was not known before.
The removal of known graphs from the output subset is motivated by the goal of exploring the underlying network of derivations.

\begin{figure}
\centering
\incFig{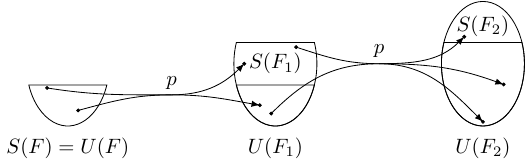}
\caption[Evaluation of Strategy Sequence]
{Illustration of the evaluation of $F_2 = p(F_1)$ for $F_1 = p(F)$ and some set of graphs, $S(F) = U(F) = \mathfrak{U}$.
Each derivation must use at least one graph form the input subset.
Two abstract derivations are shown with the endpoints indicating in which sets the graphs are.}
\label{fig:doubleRuleGraphSets}
\end{figure}

With the definition above we rewrite our initial example as; find $F_2 = p(F_1)$ for $F_1 = p(F)$ and $S(F) = U(F) = \mathfrak{U}$.
The application $p(F_1)$ can now only discover derivations with at least one graph from $S(F_1)$, which by definition contains only new graphs.
Therefore, only new derivations are found.
Fig.~\ref{fig:doubleRuleGraphSets} contains a visualization of the example.

The implementation utilizes the algorithm for transformation by first partially applying the rule to the subset of the input state, and then afterwards the full
 universe.

\section{Strategies}
The previous section described how a rule $p$ is applied to a state $F$ to calculate a new state $F'$, and motivated this by the example of composition of rule application, $F' = p(p(F))$.
Using the definition of a graph state, we generalize the interface for rule application into general strategies.
A strategy is simply any function $Q$ from and to the set of graph states.

In the following we introduce core strategies defined in the framework.
Most of the strategies are parameterized, which we will note with brackets around these parameters.
The application of a strategy $Q$ with some fixed parameter, $n$, to a graph state $F$ is thus denoted as $Q[n](F)$.

\subsection{Parallel}
A parallel strategy is defined in terms of a set of substrategies, $\{Q_1, Q_2,\dots, Q_n\}$.
The result of applying a parallel strategy is the union of the results from applying the individual substrategies:
\begin{align*}
	F' &= \stratParallel[\{Q_1, Q_2,\dots, Q_n\}](F)		\\
	U(F') &= \bigcup_{1\leq i\leq n} U(Q_i(F))		\\
	S(F') &= \bigcup_{1\leq i\leq n} S(Q_i(F))
\end{align*}

\begin{figure}
\centering
\newcommand\obabelScale{0.4}
\subfloat[]{
\incFig{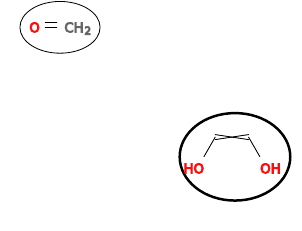}
\label{fig:formoseParallelBefore}
}
\qquad\qquad
\subfloat[]{
\incFig{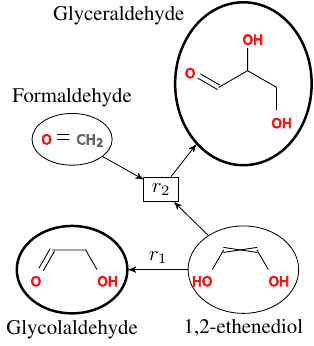}
\label{fig:formoseParallelAfter}
}
\caption[Example of a parallel strategy]{
Application of a parallel strategy $Q = \stratParallel[\{r_1, r_2\}]$ to a graph state $F$, with $r_1$ being the transformation rule for the enol to keto conversion and $r_2$ being the transformation rule for aldol addition (see Appendix~\ref{sec:formoseGrammar}).
\subref{fig:formoseParallelBefore} the reaction network with the graph state $F$ consisting of $U(F) = \{\text{formaldehyde}, \text{1,2-ethenediol}\}$ and $S(F) = \{\text{1,2-ethenediol}\}$.
\subref{fig:formoseParallelAfter} the reaction network after evaluation of $Q(F)$, with two new molecules; glycolaldehyde and glyceraldehyde.
The resulting graph state $F'$ has
$U(F') = \{\text{\text{formaldehyde}, \text{1,2-ethenediol}, \text{glycolaldehyde}, \text{glyceraldehyde}}\}$ and
$S(F') =\{\text{glycolaldehyde}, \text{glyceraldehyde}\}$.
In both networks the subset of the graph state is highlighted.
}
\label{fig:formoseParallel}
\end{figure}

\JLA{A simple use of parallel strategies is to model the possibility of   different reaction mechanisms happening simultaneously.
As example, consider modeling the formose chemistry which consists of keto-enol tautomerism and aldol addition, both reversible reactions (see Appendix~\ref{sec:formoseGrammar} for the grammar details).
Let $r_1$ and $r_2$ denote the corresponding reactions from Appendix~\ref{sec:formoseGrammar}, i.e., the enol-to-keto reaction pattern for the carbonyl group and the pattern for aldol addition.
The parallel strategy $Q =\stratParallel[\{r_1, r_2\}]$ thus models that these two reactions can happen simultaneously as illustrated in Fig.~\ref{fig:formoseParallel}.
}

\subsection{Sequence}
A sequence strategy, $Q$, is a composition of a list of substrategies, $Q_1, Q_2, \dots, Q_n$:
\begin{align*}
	Q(F) = Q_n(\dots (Q_2(Q_1(F))\dots)
\end{align*}
To increase left-to-right readability of sequence strategies, we will use the notation $Q = Q_1\rightarrow Q_2\rightarrow\dots\rightarrow Q_n$.
Additionally, if $Q_1 = Q_2 = \dots = Q_n = Q'$, we may use the normal notation for powers of functions, $Q = Q'^n$, for the sequence.

\begin{figure}
\centering
\newcommand\obabelScale{0.4}
\subfloat[]{
\incFig{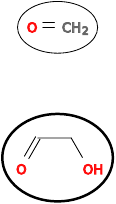}
\label{fig:formoseSequence0}
}
\hfill
\subfloat[]{
\incFig{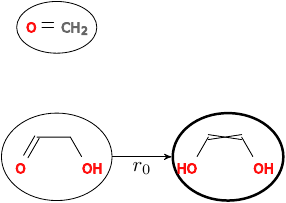}
\label{fig:formoseSequence1}
}
\hfill
\subfloat[]{
\incFig{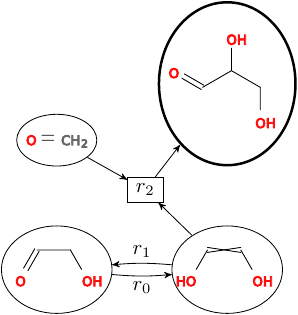}
\label{fig:formoseSequence2}
}
\caption[Example of a sequence strategy] {
Application of the sequence strategy $Q = \stratParallel[\{r_0, r_1, r_2, r_3\}]\rightarrow \stratParallel[\{r_0, r_1, r_2, r_3\}]$ to the graph state $F_0$,
with $r_i$ denoting the transformation rules for keto-enol tautomerism and reversible aldol addition.
\subref{fig:formoseSequence0} the initial reaction network with $F_0$ in which
$U(F_0) = \{\text{formaldehyde}, \text{glycolaldehyde}\}$ and $S(F_0) = \{\text{glycolaldehyde}\}$.
\subref{fig:formoseSequence1} the intermediary reaction network after evaluation of the first step of the strategy.
The difference in graph state is that 1,2-ethenediol is now added to the universe and subset, while glycolaldehyde no longer is in the subset.
\subref{fig:formoseSequence2} the reaction network after complete evaluation of $Q(F_0)$.
The final graph state $F_2$ has all four molecules in the universe and only glyceraldehyde in the subset.
Note that in the last step of the strategy the reverse keto-enol reaction is discovered, but glycolaldehyde is already in the universe so it will not be added to the subset of $F_2$.
The subset of the graph state is highlighted in each network.
}
\label{fig:formoseSequence}
\end{figure}

\JLA{An example of the application of a sequence strategy can be seen in Fig.~\ref{fig:formoseSequence}, in which two sequential steps of the formose chemistry (parallel strategies) are derived starting from a graph state $F$ with $U(F) = \{\text{formaldehyde}, \text{glycolaldehyde}\}$ and $S(F) = \{\text{glycolaldehyde}\}$.
}

\subsection{Repetition}
The sequencing strategy only allows composition of a fixed number of strategies, whereas the repetition strategy is used to compose a single strategy with itself many times.

A repetition strategy, $Q$, is parameterized by a non-negative integer, $n$, and an inner strategy $Q'$.
The inner strategy is composed with itself until the graph state reaches a fixed point or its subset is empty, however at most $n$ times:
\begin{align*}
	Q &= \stratRepeat[Q', n] = Q'^k	\\
	k &= \min \{0, 1,\dots , n\}, \text{ such that } Q'^k(F) = Q'^{k+1}(F) \vee S(Q'^{k+1}(F)) = \emptyset \vee k = n
\end{align*}
This means that if the graph state reaches a fixed point then that graph state is returned, and if the subset of the state becomes empty then the \emph{previous} state is returned.
We motivate this condition of a non-empty subset of a  produced graph state by our definition of rule application, which requires at least one graph from the subset.
By returning the last graph state with non-empty subset the repetition strategy can be used as a precomputation in a sequence to find a kind of closure under some inner strategy.

Note that \JLA{if $k=0$} the strategy becomes the identity strategy, i.e., the resulting graph state is the same as the input graph state.
If $n$ is set large enough to not limit the repetition, we call it unbounded repetition, and write it as $Q = \stratRepeat[Q']$.

\begin{figure}
\centering
\newcommand\obabelScale{0.29}
\subfloat[] {
\incFig{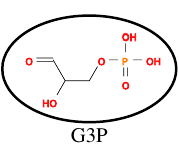}
\label{fig:g3pRepeat:0}
}
\qquad\qquad
\subfloat[] {
\incFig{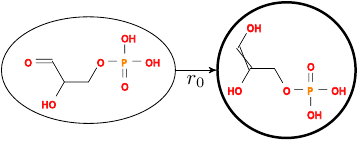}
\label{fig:g3pRepeat:1}
}
\\
\subfloat[] {
\incFig{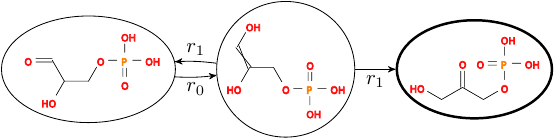}
\label{fig:g3pRepeat:2}
}
\\
\subfloat[] {
\incFig{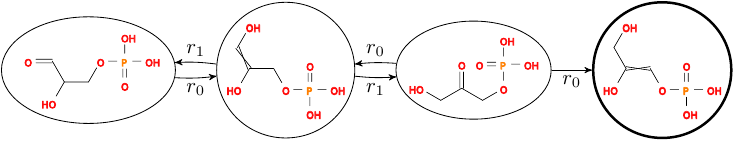}
\label{fig:g3pRepeat:3}
}
\\
\subfloat[] {
\incFig{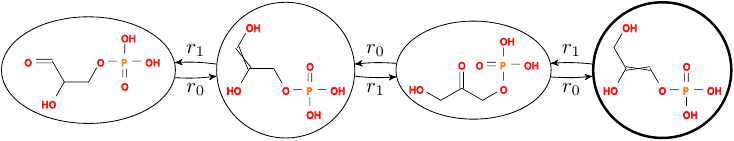}
\label{fig:g3pRepeat:4}
}
\caption[Generation of isomers using a repetition strategy] {
The strategy $Q = \stratRepeat[\stratParallel[\{r_0, r_1\}]]$ applied to the the initial graph state $F_0$ with $U(F_0) = S(F_0) = \{\text{G3P}\}$ (shown in \subref{fig:g3pRepeat:0}).
\subref{fig:g3pRepeat:1}--\subref{fig:g3pRepeat:3} the intermediary reaction networks from evaluation of $Q(F_0)$.
Each step discovers a new isomer which constitutes the new subset.
Additionally, the reaction to the previous isomer is discovered. However, this molecule is already in the universe of the current graph state and is therefore not added to the subset.
\subref{fig:g3pRepeat:4} the final step in the repetition results in an empty subset as only known molecules (those in the universe) are rediscovered.
The graph state from \subref{fig:g3pRepeat:3} is therefore the result.
In all networks the subset of the current graph state is highlighted.
}
\label{fig:g3pRepeat}
\end{figure}

\JLA{In Fig.~\ref{fig:formoseSequence} the strategy for deriving two steps of the formose network is shown.
As a generalization the strategy $Q = \stratRepeat[n, \stratParallel[\{r_0, r_1\}]]$ can be used to derive (at most) $n$ steps of the network.
Fig.~\ref{fig:g3pRepeat} shows another example using the repetition strategy, where all isomers of glyceraldehyde 3-phosphate (G3P) are generated.
}
 
\subsection{Revive}
\JLA{Consider the following high-level description of a strategy:
Given a single graph $g$, try to apply the rule $p$.
If the application of $p$ is successful, then let $H$ denote all the produced graphs and return $H\backslash \{g\}$ (all graphs not already known).
If the application of $p$ is not successful, then intentionally $\{g\}$ should be returned.
The simple strategy $Q = p$ applied to $F$ with $S(F) = U(F) = \{g\}$ only partially achieves this, as illustrated in the following.
Let $F' = Q(F)$ be the resulting graph state after evaluation of the strategy on $F$.
Using the definition of the rule application strategy, Eq.~\eqref{eq:ruleDef}--\eqref{eq:ruleSubsetDef}, we get
\begin{itemize}
\item $S(F') = H\backslash\{g\}$ and $U(F') = H\cup\{g\}$ if $p$ is successfully applied, and
\item $S(F') = \emptyset$ and $U(F') = \{g\}$ if $p$ can not be applied.
\end{itemize}
However, the desire was to have $S(F') = \{g\}$ in the unsuccessful case.
The intention of the revive strategy is to provide a mechanism to model the desired behaviour.
A rule application strategy discovers a (possibly empty) set of derivations.
We say that a graph $g$ is \emph{consumed} in a rule application strategy if any of the discovered derivations $G\Rightarrow H$ have $g\in G$.
In the natural way we extend this and say that a graph $g$ is consumed by a strategy if it is consumed by any of its substrategies.
A revive strategy, $\stratRevive[Q']$, is parameterized by a single substrategy, $Q'$, and is defined as:}
\begin{align*}
	F' &= \stratRevive[Q'](F)	\\
	U(F') &= U(Q'(F))	\\
	S(F') &= S(Q'(F)) \cup \{g\in S(F)\mid g\in U(F') \wedge	\text{$g$ is not consumed in $Q'$}	\}
\end{align*}
That is, any graph from the input subset which is still in the output universe and was not consumed, will be added to the output subset.
\JLA{The high-level described example \DM{to illustrate the problem with a simple rule $p$} can now be solved with the strategy $Q = \stratRevive[p]$.
If the application of $p$ is unsuccessful, then $g$ is not consumed and will be added to the resulting subset.}

\begin{figure}
\centering
\subfloat[$g_1$][$g_1$\phantom{$\xrightarrow{l}$}]{
\incFig{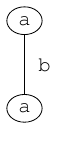}
}\qquad
\subfloat[$g_2$][$g_2$\phantom{$\xrightarrow{l}$}]{
\incFig{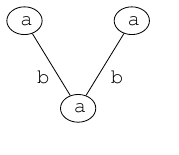}
}
\subfloat[$p = (L\xleftarrow{l}K\xrightarrow{r}R)$]{
\incFig{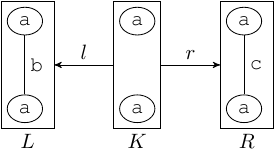}
}
\caption{Graphs and transformation rule for the example of the semantics of revive strategies.}
\label{fig:reviveEks}
\end{figure}
\begin{figure}
\centering
\incFig{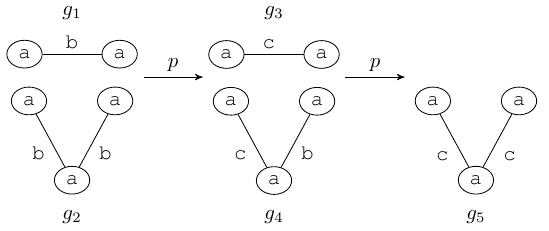}
\caption[Use Case for Revive Strategy]
{Illustration of the application of $\stratRepeat[p]$ to $F$ with $S(F) = U(F) = \{g_1, g_2\}$. Only the subset of the graph states are shown. The first application of $p$ results in two new graphs, $g_3$ and $g_4$, but as $p$ can only be applied to $g_4$ the final subset is only a single graph, $g_5$, instead of both $g_3$ and $g_5$.}
\label{fig:revive:wrong}
\end{figure}

As another example, consider the following problem. Two graphs, $g_1$ and $g_2$ and the transformation rule $p$, as illustrated in Fig.~\ref{fig:reviveEks} are given.
We wish to develop a strategy to transform all edge labels using rule $p$, with the intend to use this strategy as a precomputation for a subsequent strategy.
That is, the subset of the graph state after evaluation of the strategy must contain the completely transformed graphs in the subset.
The strategy $Q = \stratRepeat[p]$ may seem like the most intuitive approach to model this process.
However, the evaluation of $Q(F)$ with $S(F) = U(F) = \{g_1, g_2\}$ does not give the intended result, which is illustrated in Fig.~\ref{fig:revive:wrong}. 
The problem is that the repetition strategy will continue as long as \emph{any} new graph can be discovered, and does not preserve the most derived graphs in the subset.
Using the strategy $\stratRepeat[\stratRevive[p]]$ correctly solves the problem. A chemical example for the revive strategy will be given in the results section.

\subsection{Derivation Predicates}
For the purpose of precise modeling and the problems with combinatorial explosion it is convenient to limit the possibilities of expansion.
We define two variations of the concept of derivation predicates, which both introduce extra constraints in Eq.~\eqref{eq:ruleDef} to prune unwanted derivations.
\JLA{The strategy $\stratLeftPredicate[P, Q']$ is defined by the predicate $P$ on a multiset of graphs and a transformation rule, and by the substrategy $Q'$.
A candidate derivation from the graphs $G$ with the rule $p$ found by $Q'$, is only fully calculated and accepted if $P(G, p)$ is true.
A right predicate strategy, $\stratRightPredicate[P, Q']$ is also defined by a predicate and a substrategy, though with the predicate $P$ evaluating a complete derivation.
Thus, a derivation $G\xRightarrow{p} H$ is only accepted if $P(G\xRightarrow{p} H)$ is true.}

As example, given a strategy $Q'$ we wish to produce only graphs with at most 42 vertices (atoms, in a chemical context).
This requires a right predicate strategy as information about the right side of the derivation (the products) are needed.
This can be specified with the following strategy:
\begin{align*}
	Q &= \stratRightPredicate[P, Q']	\\
	P(G\xRightarrow{p}H) &\equiv \forall h\in H :  |V(h)| \leq 42
\end{align*}
\JLA{Instead, we might want to restrict that some molecule $g$ should not be an educt in any reaction with the transformation rule being $r$.
This constraint does not require the information of a complete derivation, and may as such be formulated as a left predicate strategy:
\begin{align*}
	Q &= \stratLeftPredicate[P, Q']		\\
	P(G, p) &\equiv \neg (r = p \wedge g\in G)
\end{align*}
with $Q'$ being an arbitrary strategy.}

\subsection{Filter, Sort, Take and Add}
To facilitate more elaborate use of strategies in a functional style we define several strategies which correspond to functions on lists in other languages.
As a graph state is composed of both a universe and a subset, all of these strategies are defined in two variations.

A filter strategy is parameterized by a predicate on a graph and a graph state:
\begin{multicols}{2}\noindent\small
\begin{align*}
	F' &= \stratFilterSubset[P](F)				\\
	U(F') &= U(F)							\\
	S(F') &= \{g\in S(F)\mid P(g, F)\}
\end{align*}
\begin{align*}
	F' &= \stratFilterUniverse[P](F)			\\
	U(F') &= \{g\in U(F)\mid P(g, F)\}			\\
	S(F') &= \{g\in S(F)\mid P(g, F)\}
\end{align*}
\end{multicols}

A sorting strategy is parameterized with a predicate on two graphs and a graph state, used as a less-than predicate in a stable sort of a list of graphs:
\begin{multicols}{2}\noindent\small
\begin{align*}
	F' &= \stratSortSubset[P](F)				\\
	U(F') &= U(F)							\\
	S(F') &= \stableSort[P](S(F))
\end{align*}
\begin{align*}
	F' &= \stratSortUniverse[P](F)			\\
	U(F') &=  \stableSort[P](U(F))				\\
	S(F') &= S(F)
\end{align*}
\end{multicols}
\noindent
The choice that the sorting algorithm must be stable is motivated by the desire to allow lexicographical sorting by sequencing several sorting strategies.

A take strategy is parameterized with a natural number:
\begin{multicols}{2}\noindent\small
\begin{align*}
	F' &= \stratTakeSubset[n](F)								\\
	k &= \min\{n, |S(F)|\}										\\
	U(F') &= U(F)											\\
	S(F') &= \{S(F)_1, S(F)_2, \dots, S(F)_k\}
\end{align*}
\begin{align*}
	F' &= \stratTakeUniverse[n](F)							\\
	k &= \min\{n, |U(F)|\}										\\
	U(F') &=  \{U(F)_1, U(F)_2, \dots, U(F)_k\}					\\
	S(F') &= S(F)\cap U(F')
\end{align*}
\end{multicols}

An addition strategy appends a given set of graphs to either the universe and optionally also to the subset:
\begin{multicols}{2}\noindent\small
\begin{align*}
	F' &= \stratAddSubset[\{g_1, g_2, \dots, g_n\}](F)				\\
	U(F') &= U(F) \cup \{g_1, g_2, \dots, g_n\}						\\
	S(F') &= S(F) \cup \{g_1, g_2, \dots, g_n\}
\end{align*}
\begin{align*}
	F' &= \stratAddUniverse[\{g_1, g_2, \dots, g_n\}](F)				\\
	U(F') &= U(F) \cup \{g_1, g_2, \dots, g_n\}							\\
	S(F') &= S(F)
\end{align*}
\end{multicols}

An example usage of these strategies is the procedure of ranking graphs according to some property, take the best $n$ graphs for subsequence expansion, i.e:
\begin{align*}
	Q' &= \stratSortSubset[P] \rightarrow \stratTakeSubset[n]
\end{align*}
\JLA{Note that the sorting predicate $P$ can be based on any external data such as results from wet lab experiments.
As example we have used mass spectrometry data to bias the expansion towards high intensity molecules (see the Results and Discussion section).
}

The addition strategies can be used both for injecting new graphs in the middle of a strategy, but we also find them convenient simply for uniform left-to-right writing of a strategy application.
E.g., given a (large) strategy $Q$ we wish to apply to the graph state $F$, we can write:
\begin{align*}
	F' &:= \stratAddUniverse[U(F)]\rightarrow \stratAddSubset[S(F)] \rightarrow Q
\end{align*}
with the interpretation $F' = Q(F)$.

\subsection{Implementation Remarks}
The strategies are implemented in C++ as part of a library, to allow easy extension at the user level.
Extensions can vary from simple graph state manipulating strategies to complete replacement of the underlying transformation formalism.
The library is aimed at chemical graph transformation, with special optimization for molecules (e.g., use of canonical SMILES strings for graph isomorphism~\cite{smiles, canonSmiles}),
but is not restricted to the domain of chemistry.
The current implementation uses VF2\cite{vf2} to find subgraph isomorphisms, and as a fall-back algorithm for isomorphism check for general graphs.
Furthermore, the library utilizes data structures and procedures for molecule handling form the Graph Grammar Library (GGL)~\cite{ggl}. A Python module with bindings to the C++ library is also implemented to allow easy development of expansion strategies.

\section{Results}

\DM{In this section we will apply our strategy framework to three different chemical systems and present results on how to systematically explore complex chemical universes: 
i.) for the Diels-Alder reaction system we will repeatedly merge molecules with isoprene, 
ii.) we will compare chemical universes of basic formose chemistry with and without using borate as inhibitor motivated by a recent experiment by \cite{Ricardo:2004}, and 
iii.) we will present a strategy to explore the complex chemical spaces of hydrogen cyanide polymerization 
and hydrolysis product in order to show how to integrate mass spectrometry results in our framework.  
In order to easily illustrate subspaces that are also expected to exist in a chemical setting, we will apply the strategy framework to a small puzzle game (\JLA{Appendix~\ref{sec:catalan}}).}

\subsection{The Diels-Alder Reaction}
\begin{figure}
\centering
\newcommand{\dielsAlderMolsScale}{0.30}
\subfloat[Isoprene]{
\incFig{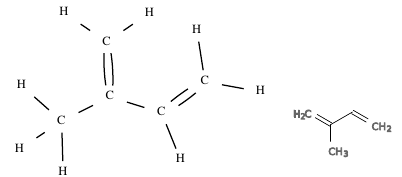}
\label{fig:isoprene}
}
\subfloat[Cyclohexadine]{
\incFig{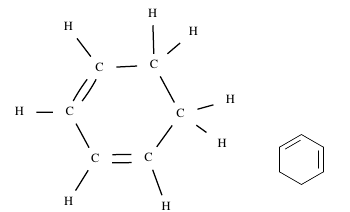}
\label{fig:cyclohexadine}
}
\caption[Starting Molecules for Chemical Space]
{The starting molecules, \subref{fig:isoprene} isoprene and \subref{fig:cyclohexadine} cyclohexadine, for application of the Diels-Alder reaction.
The molecules are shown in two versions; one with all vertices explicit and chemical interpretation of edge labels (left), and one version in standard chemical visualization.
}
\label{fig:dielsAlderStartMols}
\end{figure}

The Diels-Alder reaction is one of the most useful reactions in organic chemistry and has heavily influenced total synthesis in the last decades \cite{Nicolaou02}. The explosion of the chemical space by applying this reaction several times will be biased by the strategy framework. The reaction is shown in an example derivation in Fig.~\ref{fig:dpoDielsAlder}, while the starting molecules, isoprene and cyclohexadine, are shown in Fig.~\ref{fig:dielsAlderStartMols}.
Let $p = (L\leftarrow K\rightarrow R)$ be the transformation rule modeling the Diels-Alder reaction.
The intention of the rule is that it is applied to two molecules, but this constraint is not encoded in the rule.
We therefore first wrap $p$ with a derivation predicate:
\begin{align*}
	Q_p &= \stratLeftPredicate[P, p]		&	P(G, p') &\equiv \#G = 2
\end{align*}
This means that all derivations $G\xRightarrow{p} H$ must have $|G| = 2$.
  
A generic breadth-first exploration of the chemical space can be done with the following strategy:
\begin{align*}
	Q_{\mathrm{BFS}} &= \stratAddSubset[\{\mathrm{isoprene}, \mathrm{cyclohexadine}\}] 
		\rightarrow \stratRepeat[Q_p, n]
\end{align*} 
However, for $n = 4$ the strategy already discovers 825 new graphs through 1278 
	derivations.\footnote{In this scenario we regard derivations which only differ in the matching morphism as duplicates. The evaluation of the strategy takes in the order of 10 seconds with a \CPU.}
The number of subgraph isomorphism queries throughout the evaluation is 74591.
In Appendix~\ref{sec:dielsAlder}, Fig.~\ref{fig:dielsAlderRepeat2} the resulting derivation graph for just $n =2$ is shown.

\begin{figure}
\incFig{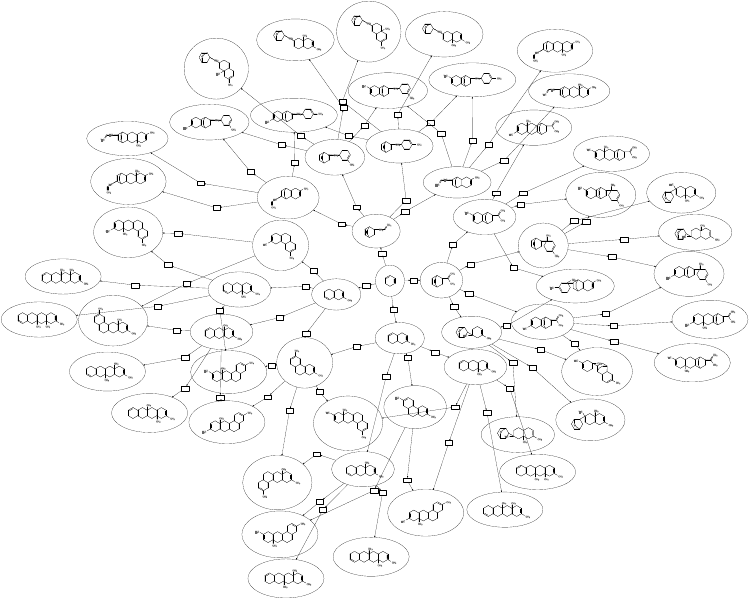}
\caption[Reaction Network Expanded by Strategy]
{The derivation graph resulting from evaluating the expansion strategy $Q_{\mathrm{subspace}}$, Eq.~\eqref{eq:dielsAlderSubspaceStrat}.
	To minimize clutter, the vertex with isoprene and the corresponding edges are not shown, although isoprene is involved in any reaction (the resulting chemical reaction network is a hypergraph).}
\label{fig:dielsAlderBias3}
\end{figure}

We now decide to only look at the subspace of molecules which are derived by repeatedly merging molecules with isoprene, starting with cyclohexadine.
The following strategy implements this specification:
\begin{align}
Q_{\mathrm{subspace}} &= \stratAddUniverse[\{\mathrm{isoprene}\}] \rightarrow \stratAddSubset[\{\mathrm{cyclohexadine}\}] 			\label{eq:dielsAlderSubspaceStrat}				\\
	&\rightarrow \stratLeftPredicate[P_{\mathrm{init}}, Q_p]								
	\rightarrow \stratFilterUniverse[P_{\mathrm{filter}}]									\notag			\\
	&\rightarrow \stratRepeat[Q_p, n]													\notag
\end{align}
with
\begin{align*}
P_{\mathrm{init}}(G, p') &\equiv G = \{\mathrm{isoprene}, \mathrm{cyclohexadine}\}		\notag			\\
P_{\mathrm{filter}}(g, F) &\equiv g\neq \mathrm{cyclohexadine}							\notag
\end{align*}
This first computes all possible proper derivations $\{\mathrm{isoprene}, \mathrm{cyclohexadine}\}\xRightarrow{p} H$, then removes cyclohexadine from the graph state to prevent further derivations.
In the end it uses breadth-first expansion for at most $n$ steps.
This strategy, with $n=3$ (i.e., 4 expansion steps including the very specific first step) discovers only 165 new graphs
	through 236 derivations,\footnote{In this scenario we regard derivations which only differ in the matching morphism as duplicates. The evaluation of the strategy takes in the order of 8 seconds with a \CPU.} and uses 5524 subgraph isomorphism queries.
The derivation graph with $n=2$ is visualized in Fig.~\ref{fig:dielsAlderBias3}.

\subsection{Borate stabilized Formose Reaction}
\newcommand\stratIndent{\quad}

Sugars, or more general carbohydrates, a broad class of organic compounds, can be viewed as polymers of formaldehyde units.
The reactivity of carbohydrates is dominated by their carbonyl and their vicinal alcohol functional groups.
In particular the enolized form of a carbonyl group may attack another one (in keto form), resulting in the formation of a new carbon-carbon bond.
This reaction is known as \textit{aldol addition} (see Fig.~\ref{fig:formoseSequence2}).
If the carbon atom adjacent to a carbonyl group carries an alcohol functionality, than the \textit{enolization reaction} of the carbonyl group erases the ``information'' at which carbon atom the carbonyl functionality was located before the enolization.
This effect allows the carbonyl group to ``travel'' along the carbohydrate backbone (see Fig.~\ref{fig:g3pRepeat:4}).
Both reactions are responsible for the meta-stability of carbohydrates and result in complex carbohydrate mixtures when repeated again and again as for instance under the conditions of the formose reaction \cite{Decker:1982}. The formose reaction has been extensively discussed as a possible prebiotic route to higher carbohydrates in particular five-carbon sugars, such as ribose, needed for the formation of nucleotides (the building blocks of RNA) \cite{Benner:2012}. Unfortunately, if the formose reaction is not stopped in time the reaction mixture turns into black ``tar''. Therefore, some stabilizing mechanism compatible with prebiotic environments, that prevent the destruction of interesting sugars, is indispensable to keep the formose reaction as a plausible prebiotic scenario for higher carbohydrate formation. The addition of borate, capable of binding vicinal diols, to the reaction mixture has been identified as such a stabilizing mechanism, that biases the outcome of the formose reaction towards high yields of five-carbon sugars \cite{Ricardo:2004}. In the following we illustrate how expansion strategies can be exploited to carve out the differences between the formose reaction networks with and without borate.

\begin{figure}
\centering
\subfloat[Formaldehyde] {
\incFig{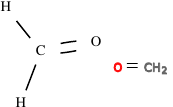}
\label{fig:formoseStartingMols:formaldehyde}
}
\subfloat[Glycolaldehyde] {
\incFig{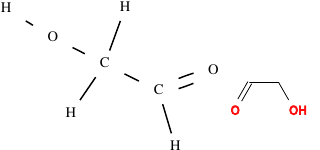}
\label{fig:formoseStartingMols:glycolaldehyde}
}
\subfloat[Borate] {
\incFig{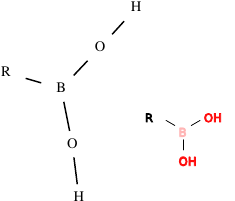}
\label{fig:formoseStartingMols:borate}
}

\subfloat[Borate $+$ 1,2-diol reaction pattern, `addBorate'] {%
\incFig{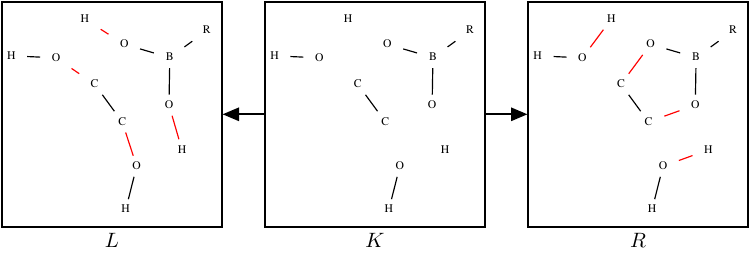}
	\label{fig:formoseStartingMols:borateReaction}
}

\subfloat[Relabeling of hydrogen to make it non-reactive, `hToD']{%
\incFig{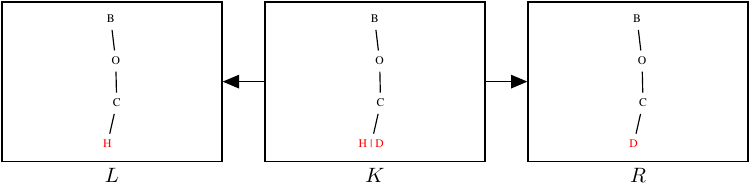}
\label{fig:formoseStartingMols:hToD}
}
\caption[The starting molecules of the formose chemistry]{
\subref{fig:formoseStartingMols:formaldehyde}--\subref{fig:formoseStartingMols:borate} the starting molecules of the borate inhibited formose reaction.
The three molecules are shown both as explicit graphs with all vertices and in standard chemical visualization.
The borate molecule \subref{fig:formoseStartingMols:borate} is modeled with only two hydroxyl groups to simplify the model.
\subref{fig:formoseStartingMols:borateReaction} the reaction pattern for forming borate complexes with 1,2-diols.
\JLA{This rule additionally has a matching constraint: none of the carbon atoms may be an endpoint of a double bond.}
To approximate the subsequent non-reactivity of the hydrogens on the carbon atoms we relabel them to \texttt{D} using the reaction `hToD' \subref{fig:formoseStartingMols:hToD}.
This relabeling is in the context graph, $K$, represented with the annotation \texttt{H | D}.
}
\label{fig:formoseStartingMols}
\end{figure}

The basic formose reaction consists of two types of reversible reaction patterns, keto-enol tautomerism and aldol reaction.
As they are reversible they are modeled by two transformation rules each.
These are shown in Appendix~\ref{sec:formoseGrammar} as transformation rule $r_0, \dots, r_3$, while the two initial molecules, formaldehyde and glycolaldehyde, are shown in Fig.~\ref{fig:formoseStartingMols:formaldehyde} and~\ref{fig:formoseStartingMols:glycolaldehyde} respectively.
To keep the model simple we use a borate-like molecule, Fig.~\ref{fig:formoseStartingMols:borate}, with just two hydroxyl groups instead of a complete molecule.
To enable the formation of borate complexes we use the transformation rule shown in Fig.~\ref{fig:formoseStartingMols:borateReaction}.
This reaction pattern is described in \cite{Benner:2010} as inhibiting keto-enol tatutomerism by making the hydrogen atoms attached to the carbon atoms non-acidic.
To approximate this behaviour we relabel these vertices from \texttt{H} to \texttt{D}, thereby preventing the reaction pattern of enolization ($r_0$ in Appendix~\ref{sec:formoseGrammar}) from matching at these locations.
The relabeling is done with the reaction `hToD', Fig.~\ref{fig:formoseStartingMols:hToD}.

The formose chemistry contains an infinite number of molecules, so to limit the scope of the exploration we prune any reaction which creates molecules with more than 5 carbon atoms.
This is formulated with a right predicate strategy around the application of the basic formose reaction patterns:
\begin{align*}
	 &\stratRightPredicate[P_{\#C}, \stratParallel[\{r_0, r_1, r_2, r_3\}]]						\\
	P_{\#C}(G\xRightarrow{p}H) \equiv& \forall h\in H : \text{$h$ has at most 5 carbon atoms}
\end{align*}
As a reference, we generate the non-inhibited reaction network with the strategy $Q_{\mathrm{BFS}}$:
\begin{align*}
Q_{\mathrm{BFS}} =& \stratAddUniverse[\{\text{formaldehyde}\}]		\\
\rightarrow	& \stratAddSubset[\{\text{glycolaldehyde}\}]					\\
\rightarrow & \stratRepeat[													\\
	&\stratIndent\phantom{\rightarrow}\ \stratRightPredicate[P_{\#C}, \stratParallel[\{r_0, r_1, r_2, r_3\}]]	\\
	& ]																		
\end{align*}
Not all molecules can actually bind with borate and must therefore be preserved while the other molecules form complexes.
This is modeled with a revive strategy around the actual complex forming reaction pattern, `addBorate'.
After the potential forming of a borate complex, the relevant hydrogen atoms must be made inactive using the rule `hToD'.
The number of relevant hydrogens may not be the same for alle molecule and therefore the relabeling strategy is embedded in both a repeat and revive strategy.
This models the notion of ``as many times as possible'' on a collection of molecules.
The reaction network with borate inhibition can thus be calculated by the following strategy:
\begin{align*}
Q_{\mathrm{borate}} =& \stratAddUniverse[\{\text{formaldehyde}, \text{borate}\}]	 	\\
\rightarrow	& \stratAddSubset[\{\text{glycolaldehyde}\}]					\\
\rightarrow & \stratRepeat[													\\
	&\stratIndent\phantom{\rightarrow}\ \stratRevive[\mathrm{addBorate}]			\\
	&\stratIndent \rightarrow \stratRepeat[\stratRevive[\text{hToD}]]		\\
	&\stratIndent \rightarrow \stratRightPredicate[P_{\#C}, \stratParallel[\{r_0, r_1, r_2, r_3\}]]	\\
	& ]	
\end{align*}
Let $\mathcal{G}$ denote the set of molecules used and generated by the evaluation of $Q_{\mathrm{borate}}$ on the empty graph state.
This set of molecules contain both borate complexes and simple carbohydrates without boron.
To canonicalize the molecules we can use the strategy
\begin{align*}
Q_{\mathrm{canon}} = \stratAddSubset[\mathcal{G}]
	&\rightarrow \stratRepeat[\stratRevive[\mathrm{dToH}]]\\
	&\rightarrow \stratRepeat[\stratRevive[\mathrm{removeBorate}]]
\end{align*}
with `removeBorate' being the inverse transformation rule of `addBorate', and `dToH' being the inverse of `hToD'.
Note that `removeBorate' requires water molecules as educts, but if `addBorate' was ever used in $Q_{\mathrm{borate}}$ these molecules must be in $\mathcal{G}$.

As a variant of the network, we also calculate the network with a an extra molecule, dihydroxyacetone, in the subset:
\begin{align*}
Q_{\mathrm{borate}}^+ =& \stratAddUniverse[\{\text{formaldehyde}, \text{borate}\}]	 	\\
\rightarrow	& \stratAddSubset[\{\text{glycolaldehyde}, \text{dihydroxyacetone}\}]					\\
\rightarrow & \stratRepeat[													\\
	&\stratIndent\phantom{\rightarrow}\ \stratRevive[\mathrm{addBorate}]			\\
	&\stratIndent \rightarrow \stratRepeat[\stratRevive[\text{hToD}]]		\\
	&\stratIndent \rightarrow \stratRightPredicate[P_{\#C}, \stratParallel[\{r_0, r_1, r_2, r_3\}]]	\\
	& ]	
\end{align*}

\begin{figure}
\centering
\incFig{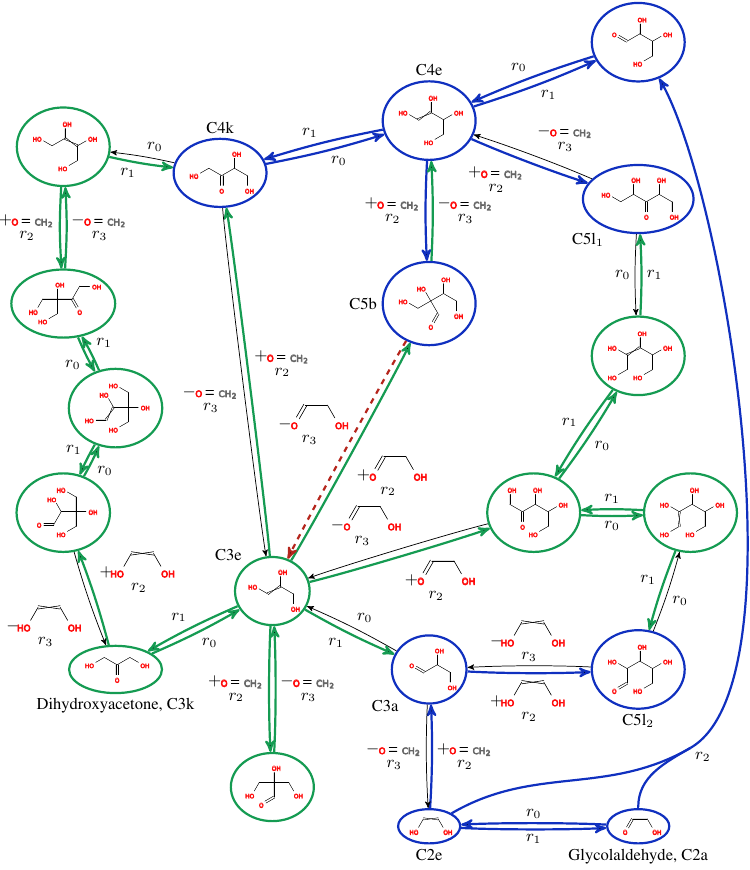}
\caption[Borate inhibited formose network] {
The reaction network of the formose chemistry as calculated with the strategy $Q_{\mathrm{BFS}}$.
The blue subnetwork correspond to the borate inhibited network calculated with $Q_{\text{borate}}$.
The green and blue networks together \JLA{with the red reaction (C5b to C3e)} correspond the network calculated with $Q^+_{\text{borate}}$, i.e., with dihydroxyacetone as an input compound.
Note that this particular model does not include stereochemical properties,
and that the molecule depictions are made using Open Babel~\cite{obabel}, which for instance means crossing double are used to indicate the unspecified stereo.
Each reaction is annotated with the reaction pattern, $r_i$, used to realize the concrete reaction.
For the aldol reactions, $r_2$ and $r_3$, the secondary educt ($+$) or product ($-$) is additionally shown.
\JLA{The addition of borate in $Q^+_\text{borate}$ is done with the strategy $\stratRevive[addBorate]$, meaning that at most 1 borate is added in each iteration.
The red reaction is no longer available if the addition is done with the strategy $\stratRepeat[\stratRevive[addBorate]]$, meaning ``add as many as possible''.}
}
\label{fig:formoseBoron}
\end{figure}
In Fig.~\ref{fig:formoseBoron} the reference reaction network created with $Q_{\mathrm{BFS}}$ is shown. Reactions in black are active only in the basic formose reaction case with formaldehyde and glycolaldehyde as set of input molecules. If borate is added to the input set of molecules, the reactions highlighted in blue are active, while the rest of the network is inactive. Finally if dihydroxyacetone is added to the input set of molecules the reactions highlighted in green are activated in addition to the blue part of the network.
The evaluation of $Q_{\mathrm{borate}}$ leaves only the blue reactions, which are selective pathways from glycolaldehyde (C2a) to five-carbon sugars (C5b, C5l$_1$, C5l$_2$) active, while the rest of the network is shut down via borate inhibition. These pathways rely on a constant replenishment of glycolaldehyde. Here dihydroxyacetone (C3k) comes into play. C3k can only be formed from within the formose network via retro-aldol reaction from higher carbohydrates. If added to the reaction network an catalytic loop is activated (sub-network in green: C3k, C3e, C4k, C4e, C5b, retro-aldol red dashed arrow to C3e and C2a) supporting the blue sub-network since C2a ends up as some five-carbon sugars in the blue sub-network. C3e enters another round in the cycle to construct another C2a. These computational results are in very good agreement with the experimental results presented in \cite{Kim:2011}.

\subsection{HCN Polymerization and Hydrolysis biased by Mass Spectrometry Results}

Hydrogen cyanide (HCN) is a known prebiotic precursor of amino acids as well as many other molecules relevant to present-day biology. 
It has been used to synthesize adenine already in 1961 \cite{Oro61}  amino acids \cite{Ferris74}, as well as many other molecules relevant to present-day biology \cite{Ferris74,Ferris78,Voet83,Miyakawa02,Saladino:04,Borquez:05,Matthews:06}, 
and it is also known to play a key role also in sugar synthesis \cite{Ritson12}.
In \cite{ownHCN} graph grammar approaches and mass spectrometry results were integrated in order to generate a chemical network with highly likely polymerization/hydrolysis products. 
In the first step of the wetlab experiments acid-catalyzed HCN polymers were created, in the second step the polymers were hydrolysed under different conditions. 
The mass spectrometry results of the wetlab experiments were used in order to bias the chemical space exploration performed with the strategy framework. 
A detailed discussion of the results including a large variety of adenine pathways and autocatalytic processes within the inferred chemical space can be found in \cite{ownHCN}. 
Here we focus on the description of the used strategies.

The model of the HCN chemistry is based on many transformation rules which are shown in detail in the web supplement of \cite{ownHCN}.
For the purpose of a concise strategy description we let $\mathfrak{R}$ denote the set of needed transformation rules.
The expansion strategy is aimed at modeling the wetlab experiments and thus consist of the sequencing of a strategy for polymerization with a strategy for hydrolysis.
As these two strategies are quite similar we only state the hydrolysis strategy, $Q_{\text{hydrolysis}}$.

Ideally, a simple breadth-first expansion strategy, $\stratRepeat[\mathfrak{R}]$, can be used to expand the network but due to the sheer combinatorial explosion only very few steps can be calculated.
Instead the following strategy can be used to prune the expansion:
\begin{align*}
Q_{\mathrm{hydrolysis}} =& \stratAddSubset[\{\mathrm{HCN}, \mathrm{NH}_4, \mathrm{H}_2\mathrm{O}, \mathrm{OH}^-\}]								\\
\rightarrow & \stratRepeat[														\\
	&\stratIndent\phantom{\rightarrow\ }\stratLeftPredicate[P, \stratParallel[\mathfrak{R}]]	\\
	&\stratIndent\rightarrow \stratFilterUniverse[P_{\mathrm{isomer}}]		\\
	&\stratIndent\rightarrow \stratSortUniverse[P_{\mathrm{intensity}}]		\\
	&\stratIndent\rightarrow \stratTakeUniverse[20]								\\
	&\stratIndent\rightarrow \stratAddUniverse[G_{\text{small}}]	\\
	&]																			\\
\end{align*}
where the predicates are defined as
\begin{align*}
P(G, p)\equiv & \text{ at most 1 molecule of $G$ has molar mass greater than 50}		\\
P_{\mathrm{isomer}}(g, F) \equiv& \text{ $true$ iff the normalized Boltzmann factor of $g$ is above}	\\
	&\text{ a certain threshold.}			\\
	&\text{The factor is calculated based on the isomers of $g$ in $U(F)$}			\\
P_{\mathrm{intensity}}(g_1, g_2, F) \equiv&\text{ } intensity(g_1) > intensity(g_2)		\\
	& \text{The intensities are found in the mass spectrometry data}\\
	& \text{using the molar masses }
\end{align*}
That is, the input graph state is augmented with basic food molecules.
Then the main hydrolysis step is repeated until no new molecules are found.
The main step first expands the network under the constraint that at least one small molecule is an educt in each reaction, which limits the growth of the molecules to be linear as opposed to exponential.
The subsequent three steps prune the graph state of unlikely molecules, first by calculating normalized Boltzmann factors within each class of isomers.
Then the mass spectrometry data from the wetlab experiments are used to select the 20 molecules with highest intensity for the next expansion step.
These pruning steps might have removed the basic food molecules, and they are therefore reintroduced.
Additionally the molecules immediately derivable from the food molecules are added.
The evaluation of the overall HCN strategy take considerably longer (hours) to calculate than the previous examples.
The bulk of the time is however spent on calculating energy value used the Boltzmann factors. For further details see \cite{ownHCN}.

\section{Conclusions}
We have introduced here a generic framework to specify and execute
strategies for the systematic exploration of spaces of graphs. Our
generative approaches use the Double Pushout formalism to derive new
graphs. Since this task is of immediate practical relevance in chemistry,
we designed our framework and implementation with the aim of high
efficiency in this particular domain of application. 
As performance was a particular focus of our work, we use state-of-the-art
subgraph isomorphism check methods and we heavily employ hashing techniques
in the checks for graph isomorphism; in order to infer proper derivations
of new molecules with full or partial rule application we do \emph{not} use
a straightforward method to enumerate all possible left-hand-sides of
derivations. Instead we employ partial rule applications, a method that
shows theoretically as well as empirically a much better performance. The
latter aspect will discussed in more detail elsewhere.

As showcase examples we have considered complex systems of chemical
reactions. For Diels-Alder reactions, which is plagued by a very rapid
combinatorial explosion, we used the strategy framework to guide the
exploration to emphasize products of repeated isoprene addition instead of
unconstrained combinations of reactions products. This is of relevance
e.g.\ in terpene chemistry and biosynthesis. In the case of the formose
reaction we show how the strategies framework can be applied to explaining
the effects of additional reactants on a given reaction network. In
particular, we can in rule based manner also determine which reactions are
effectly superseeded by new ones, so that additional reactants can lead to
a reduction of chemical network. The strategies framework thus serves not
only as a convenient tool for exploration but allows also a detailed
modelling of contraints in chemical networks. 

Although the design was clearly chosen with systems chemistry and systems biology applications in mind, the strategy framework introduced here is however by no means limited to chemical
applications. Another promising application is the emulation of
higher-level rules. In the DPO graph grammar formalism, the size of a
subgraph that is affected by a transformation is by construction bounded by
the left graph of the production that is to be applied. Apparently simple
operations on a graph, such as \emph{``contract a clique in $G$ to a single
  vertex''}, however, do not have such a bound since the clique sizes
depend only on the input graph.  Hence, such rules cannot be specified
directly as productions in a DPO graph grammar. In the Appendix~\ref{sec:catalan}
we use the well-known Catalan game \cite{catalan} to show how our strategy
framework can be applied to emulate this type of higher-level rules.

In order to analyze chemical reaction networks as created by our strategy
framework, there exist several mathematical techniques that we plan to
apply to our generated networks.  Two of the most prominent ones are Flux Balance Analysis~\cite{fba}
and Elementary Mode Analysis~\cite{ema}.  Note, that these methods are
usually not applied to dynamically created reaction networks as produced by
our framework.  We aim at detecting new well-defined chemical reaction
pattern.  Furthermore, we expect to identify highly connected subgraphs in
chemical spaces, that are connected via a small number of bridging
reaction, similar to our observation for the Catalan game.

    
\section{Authors contributions}
J.L.A.\ implemented the strategy framework. All authors contributed to
the theory, the writing of the manuscript and approved the submitted
manuscript.

\section{Acknowledgements}

This work was supported in part by the Volkswagen Stiftung proj.\ no.\
I/82719, the COST-Action CM0703 ``Systems Chemistry'', and the Danish
Council for Independent Research, Natural Sciences.

\bibliographystyle{unsrt}
\bibliography{dgStrat}


\appendix
\section{Transformation Rules for the Formose Chemistry}
\label{sec:formoseGrammar}
The main formose chemistry consists of two reversible reactions, keto-enol tautomerism and aldol addition.
These reaction patterns are listed below as four transformation rules, $r_0$ to $r_3$, one for each direction.
Additionally, for modeling borate inhibition we use a borate addtion rule, $r_4$.
The inverse of this rule, $r_5$, is used for generating the underlying molecule without borate.
The rules $r_6$ and $r_7$ are used for converting between acidic and non-acidic hydrogens in borate complexes.
Note that the context graph, $K$, of $r_6$ and $r_7$ also uses the labeling scheme ``L label\texttt{ | }R label'', with the meaning that the vertex changes label from ``L label'' to ``R label''.


\subsection{$r_0$, Keto-enol Tautomerism, Keto-to-enol}
\incFig{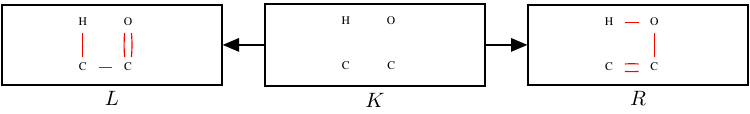}
\subsection{$r_1$, Keto-enol Tautomerism, Enol-to-keto}
\incFig{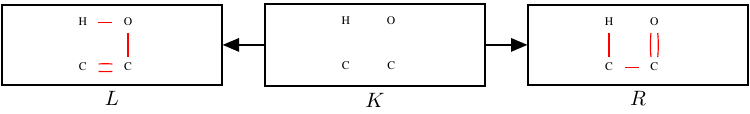}
\subsection{$r_2$, Aldol Reaction, Addition}
\incFig{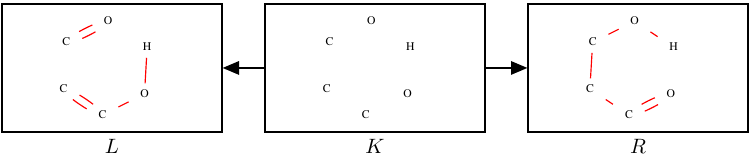}
\subsection{$r_3$, Aldol Reaction, Splitting}
\incFig{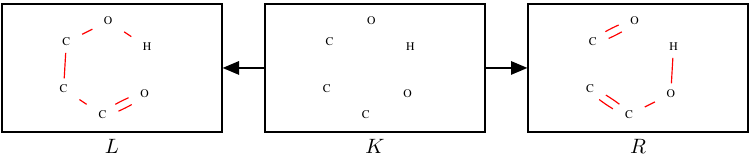}
\subsection{$r_4$, Borate Reaction, Addition}
\incFig{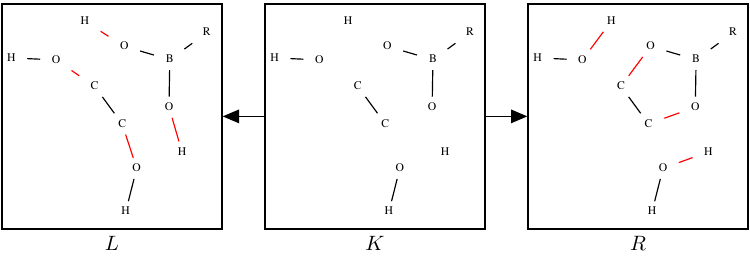}
The rule has the following matching condition: none of the adjacent edges of the carbon vertices may represent a double bond.

\subsection{$r_5$, Borate Reaction, Splitting}
\incFig{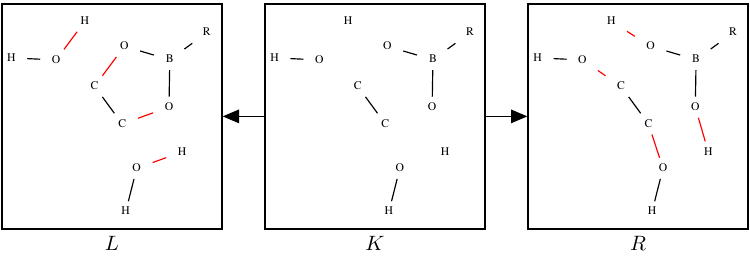}
\subsection{$r_6$, Acidic to Non-acidic Hydrogen}
\incFig{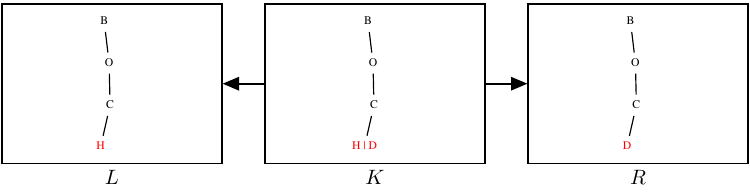}
\subsection{$r_7$, Non-acidic to Acidic Hydrogen}
\incFig{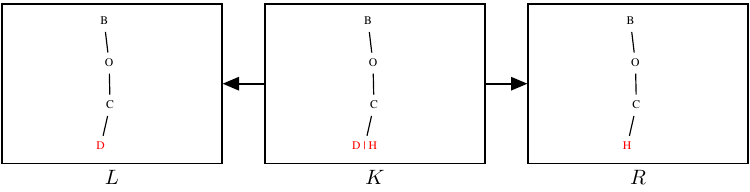}

\section{Additional Diels-Alder Chemistry Figure}
\label{sec:dielsAlder}
\begin{figure}[h]
\incFig{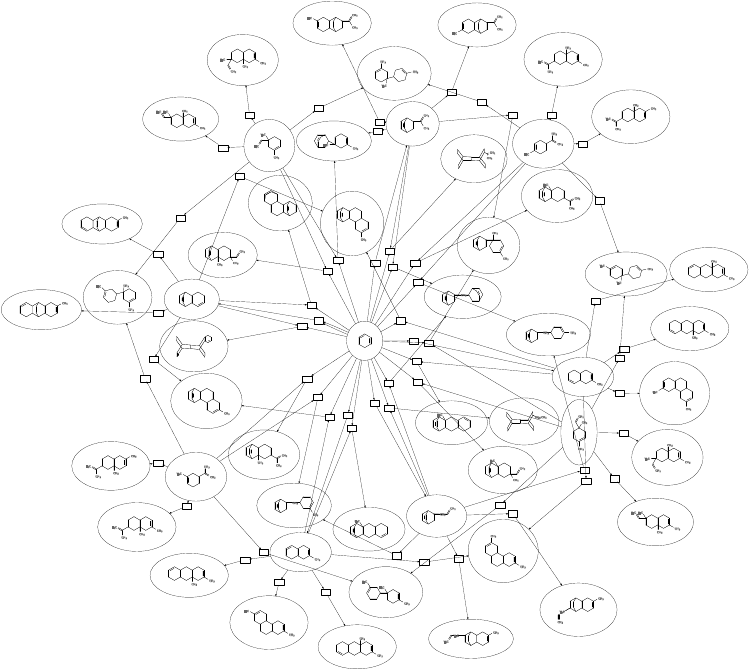}
\caption{The derivation graph resulting from evaluating the breadth-first expansion strategy
	$Q_{\mathrm{BFS}} = \stratAddSubset[\{\mathrm{isoprene}, \mathrm{cyclohexadine}\}] \rightarrow \stratRepeat[Q_p, 2]$ (on an empty graph state).
	To minimize clutter, the vertex with isoprene and the corresponding edges are not shown.}
\label{fig:dielsAlderRepeat2}
\end{figure}

Fig.~\ref{fig:dielsAlderRepeat2} shows the derivation graph obtained from the breadth-first expansion of the Diels-Alder chemistry.
The number of expansion steps is only 2.

\section{Solving the Catalan Game}
\label{sec:catalan}
\begin{figure}
\centering
\incFig{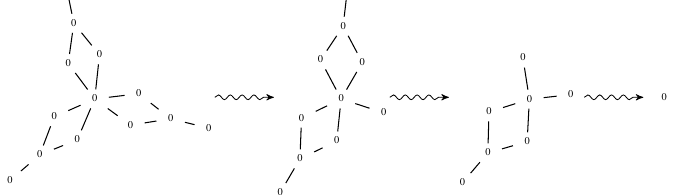}
\caption{Level 1 of the Catalan game and the intermediary graphs during transformation to a graph with a single vertex.}
\label{fig:catalanExample}
\end{figure}

The Catalan game~\cite{catalan} is a puzzle game in which the player in each level is presented with a simple undirected graph without labels.
The goal is to transform the graph into a single vertex using the following rewriting rule; given a vertex $v$ with degree exactly 3, identify $v$ with its neighbours and preserve simpleness of the graph by identifying parallel edges and deleting loops.
Fig.~\ref{fig:catalanExample} shows level 1 with the intermediary graphs towards the goal graph with a single vertex.

The transformation in the game can not be formulated as a single rule in the DPO formalism, because such rules must explicitly match the vertices and edges which are changed, while the Catalan transformation needs to change arbitrarily many edges.
In the following we show how the strategies can be used to implement a move in the game, using only DPO rules.

\begin{figure}
\centering
\subfloat[]{
\incFig{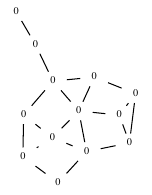}
\label{fig:catalan25}
}\qquad
\subfloat[]{
\incFig{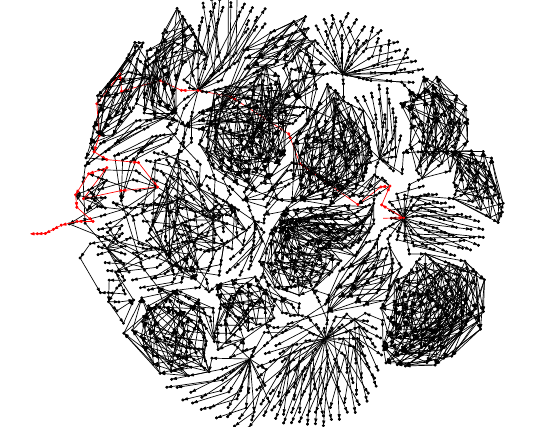}
\label{fig:catalanSpace}
}
\caption{The derivation graph created during expansion of level 25 of the Catalan game. A path equivalent to a solution is highlighted.\\}
\end{figure}

Let $g$ be the graph from some Catalan level, with all edge labels set to the empty string and all vertex labels set to the arbitrarily chosen label ``\texttt{0}''.
A high-level description of a move is:
\begin{enumerate}
\item Find a vertex $v$ with at least 3 neighbours and mark it by changing the label to ``\texttt{A}''.
	Mark the 3 matched neighbours with the label ``\texttt{R}''.
\item If possible, find another fourth neighbour of $v$ and mark $v$ with ``\texttt{FAIL}''.
\item Discard all graphs with a vertex with the label ``\texttt{FAIL}''.
\item For all edges $e$ with both end-vertices having label ``\texttt{R}'', remove $e$.
\item For all edges $ur$ with $u$ having label ``\texttt{0}'' and $r$ having label ``\texttt{R}'', add $uv$ if it does not exist already and then remove $ur$.
\item For all edges $ur$ with $u$ having label ``\texttt{0}'' and $r$ having label ``\texttt{R}'', remove $ur$.
\item Remove all neighbours of $v$ having label ``\texttt{R}''.
\item Unmark $v$ by changing the label to ``\texttt{0}''.
\end{enumerate}
Step 3 can be implemented with a filtering strategy while the other steps each require a transformation rule.
The following strategy can be used to solve a level, in the sense that if a graph with a single vertex with label ``\texttt{0}'' is found, then a path to that graph is equivalent to a solution.
The details of the transformation rules (mark, markForFail, removeInterR, reattachExternal, removeAttached, removeR and unmark) are shown in Appendix~\ref{sec:catalanRules}.
\newcommand{\ind}{\qquad}
\begin{align*}
Q_{\mathrm{catalan}}	=&\stratAddSubset[\{g\}]				\rightarrow\stratAltRuleApp[\stratRepeat[			\\
&\ind \mathrm{mark}																	
 \rightarrow\stratRevive[\mathrm{markForFail}]									
 \rightarrow\stratFilterUniverse[P_{\mathrm{fail}}]										\\
&\ind \rightarrow\stratRepeat[\stratRevive[\mathrm{removeInterR}]]					\\
&\ind \rightarrow\stratRepeat[\stratRevive[\mathrm{reattachExternal}]]					\\
&\ind \rightarrow\stratRepeat[\stratRevive[\mathrm{removeAttached}]]					\\
&\ind \rightarrow \mathrm{removeR}													
 \rightarrow \mathrm{unmark}															\\
&]]																						\\
P(g', F) \equiv &\text{ no vertex of $g'$ has the label ``\texttt{FAIL}''}
\end{align*}
With strategy $Q_{\mathrm{catalan}}$ all 56 levels of Catalan could be solved, all but one level took less than 10 minutes of computation time. Fig.~\ref{fig:catalanSpace} exemplarily shows the derivation graph created when executing the strategy with $g$ encoding level 25 of the game, and Fig.~\ref{fig:catalan25} show the initial level graph.
The resulting derivation graph is, in contrast to chemical reaction networks, not a hypergraph. 
However, the graph clearly illustrates subspaces that are connected via a small number of bridging edges.
Such subspaces are also expected in chemical reaction networks.

\section{Transformation Rules for the Catalan Game}
\label{sec:catalanRules}
The following sections contain visualization of the rules used in the strategy to solve a level in the Catalan game.
Vertices and edges shown in red are those being changed during transformation.
For some vertices the change is only a change of label. The label in the context graph, $K$, is for those in the format ``\texttt{L | R}'' with \texttt{L} and \texttt{R} being the label in the left and right side of the rule.

\subsection{mark}
\incFig{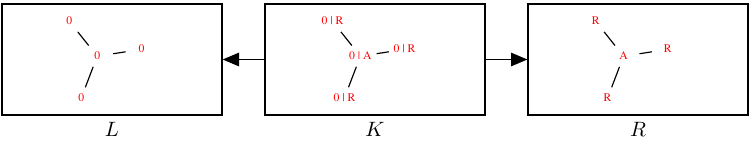}
\subsection{markForFail}
\incFig{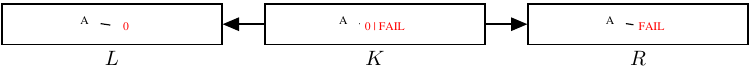}
\subsection{removeInterR}
\incFig{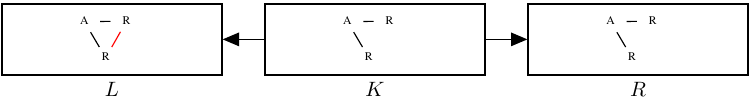}
\subsection{reattachExternal}
\incFig{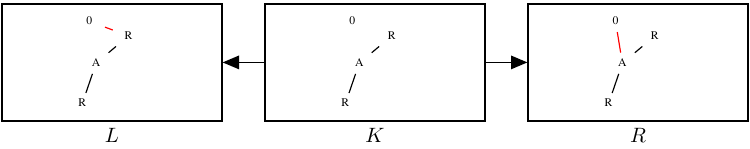}
\subsection{removeAttached}
\incFig{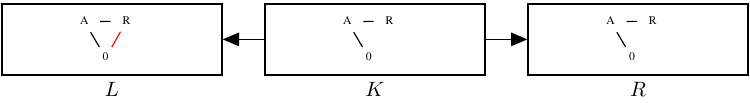}
\subsection{removeR}
\incFig{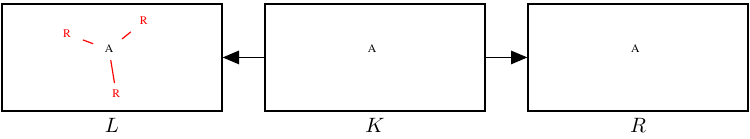}
\subsection{unmark}
\incFig{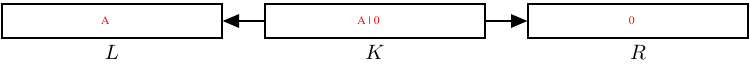}

\begingroup
\parindent 0pt
\parskip 2ex
\def\enotesize{\normalsize}
\endgroup

\end{document}